\documentclass[10pt]{article}
\usepackage[latin1]{inputenc}
\usepackage{amscd}
\usepackage{amsmath}
\usepackage{amssymb}
\usepackage{amsfonts}
\usepackage{latexsym}
\usepackage{amsthm}
\usepackage{mathrsfs}
\usepackage{booktabs}
\usepackage{bbm}
\usepackage{caption}
\usepackage[nohead]{geometry}
\usepackage{epsfig}
\usepackage[lined,boxed,commentsnumbered]{algorithm2e}
\usepackage{authblk}
\usepackage{natbib}

\numberwithin{equation}{section}

\def\0{\mbox{\bf{0}}}

\newcounter{example}[section]
\def\theexample{\thesection.\arabic{example}}
{
\medskip\noindent\\
\refstepcounter{example} {\bf Example \theexample\hspace{2pt}}
-\hspace{2pt}}{%
 \begin{flushright}
 {\footnotesize $\square$}
 \end{flushright}
}

\title{\huge On Model Selection with Summary Statistics}
\author{Erlis Ruli\\
\texttt{ruli@stat.unipd.it}}
\affil{Department of Statistical Sciences, University of Padova, Italy}

\begin{document}

\maketitle

\date{}
\begin{abstract}
Recently, many authors have cast doubts on the validity of ABC model choice. It has been shown that the use of sufficient statistic in ABC model selection leads, apart from few exceptional cases in which the sufficient statistic is also cross-model sufficient, to unreliable results. In a single model context and given a sufficient summary statistic, we show that it is possible to fully recover the posterior normalising constant, without using the likelihood function. The idea can be applied, in an approximate way, to more realistic scenarios in which the sufficient statistic is not unavailable but a ``good" summary statistic for estimation is available.
\end{abstract}

\section{Background}
Approximate Bayesian Computation (ABC) is a useful tool for Bayesian \citep[see, e.g.,][]{marin2012} or frequentist \citep[see, e.g.,][]{rubio2013simple} inferences when the likelihood function is mathematically or computationally unavailable. The successfulness of the ABC method relies on a careful choice of: the summary statistics $s(\cdot)$, the distance metric $\rho(\cdot,\cdot)$ and tolerance level $\epsilon>0$; with the summary statistic playing arguably the most crucial role.

To set the notation,  let $y = (y_1,\ldots,y_n)$ be $n$ realisations of the random variable $Y\sim P_\theta$, with $\theta\in\Theta\subseteq\mathbb{R}^d$, $d\geq1$. Furthermore, let $\pi(\theta)$ be a prior distribution for $\theta$, for simplicity assumed to be proper,  let $L(\theta)$ be the likelihood function based on model $P_\theta$ and data y and let $\pi(\theta|y)\propto L(\theta)\pi(\theta)$ be the posterior distribution, with normalising constant $p(y) = \int_{\theta\in\Theta}L(\theta)\pi(\theta)\,\mbox{d}\theta$.  The Bayes factor (BF), the standard Bayesian solution for model selection, involves the posterior normalising constants of the models under comparison. Thus, if the likelihood for a single model is unavailable, the BF cannot be computed. 

The ABC machinery comes equipped  with an ABC model choice (ABC-MC) algorithm which works as follows (\citealp{grelaud2009abc}).

\vspace{1em}
\IncMargin{1em}
\begin{algorithm}[H]
  \SetAlgoLined
  \KwResult{A sample of model indices $(j^{(1)},\ldots,j^{(N)})$}
  \For {$i = 1 \to N$}{
  \Repeat{$\rho(s(y^*), s(y))\leq\epsilon$}{
    \nl draw $j^*$ from $\pi(M=j)$\\ 
    \nl draw $\theta^*_{j^*}$ from $\pi_{j^*}(\theta_{j^*})$\\
    \nl draw $y^*$ from $p_{j^*}(\cdot;\theta^*_{j^*})$\\
}
 \nl set $j^{(i)}=j^*$
}
  \caption[]{\label{alg:abc} ABC model choice (ABC-MC) sampler.}
\end{algorithm}\DecMargin{1.em}
\vspace{1em}

where $\epsilon$ is the threshold value and $j=1,\ldots,m$ is the model index. The $N$-vector of indices $j$ produced form Algorithm~\ref{alg:abc} can be used, in principle, to compute posterior model probabilities and BFs.

Recently, many authors have cast doubts on the validity of the ABC model choice procedure \citep[see, e.g.,][]{marin2014relevant,robert2011lack}. For instance, suppose $y$ is a vector of counts and we wish to choose between the Poisson and the Geometric model. In both cases, with ABC we can obtain (almost) the exact posterior by using $\sum_{i}{y_i}$ as the summary statistic, since the latter is sufficient under both models. However, the BF obtained with ABC-MC in this case converges asymptotically to a positive constant (\citealp{robert2011lack}). With the particular exception of Gibbs random fields \citep{grelaud2009abc}, the BF obtained with ABC-MC misses the exact BF by some unknown function of the data. \cite{marin2014relevant} give theoretical conditions under which the summary statistic gives valid posterior model probabilities or BFs under the ABC model choice framework.

Clearly, the issue is with the summary statistic $s(\cdot)$. Even though it can be sufficient for the parameters, it is the cross-model sufficiency that plays the crucial role here, e.g., the summary statistic must be sufficient for the models themselves (see also \citealp{marin2014relevant}).  Finding cross-model sufficient statistics in practice is impossible, and some efforts have been spent on constructing summary statistics for ABC model selection (see, e.g., \citealp{barnes2012}). However, at the best of our knowledge, the choice of summary statistics for ABC model selection is still an open problem. Last but not the least, summary statistics for ABC model selection are notoriously a bad choice for ABC posterior sampling (C.P. Robert, personal communication). 

In Section 2 we show how the marginal likelihood can be approximated by using the sufficient summary statistic and ABC. In Section 3 we conclude by pointing to future developments.

\section{Marginal likelihood from sufficient statistics}

Let us focus on a single model $P_\theta$, and suppose that $s(\cdot)$ is sufficient for $\theta$. By the sufficiency principle we have that
$$
\pi(\theta|y) = \frac{L(\theta;y)\pi(\theta)}{p(y)} = \pi(\theta|s(y))\,.
$$

From this we see that
$$
p(y) = \frac{L(\theta;y)\pi(\theta)}{\pi(\theta|y)} =  \frac{L(\theta;y)\pi(\theta)}{\pi(\theta|s(y))}\,.
$$ 

To approximate $p(y)$ we propose to approximate $\pi(\theta|s(y))$ via ABC, and  $L(\theta;y)$ by simulation as follows (see Algorithm~\ref{alg:abcmio}).

\vspace{1em}
\IncMargin{1em}
\begin{algorithm}[H]
  \SetAlgoLined
  \KwResult{approximation of $p(s(y))$.}
  \nl use ABC to get a posterior sample and compute its mean $\hat\theta$;\\
  \nl compute $\hat{\pi}(\hat{\theta}|s(y))$, the ordinate of the posterior at $\theta = \hat\theta$;\\
   \nl draw a large sample of $y^*$ from $p(\cdot;\hat{\theta})$ and;\\
  \nl compute $\hat p(y;\hat{\theta}) = \hat L(\hat{\theta};y)$ an approximation of $p(y;\hat{\theta})$; \\
  \nl set $\hat p(y) = \frac{\hat p(y;\hat{\theta})\pi(\hat{\theta})}{\hat{\pi}(\hat{\theta}|y)}$.
  \caption[]{\label{alg:abcmio} Marginal likelihood from summary statistic.}
\end{algorithm}\DecMargin{1.em}
\vspace{1em}

Steps 2 and 4 of Algorithm~\ref{alg:abcmio} can be performed by any density estimation method; in Step 5 we only need to generate a (possibly) large sample of data from the model under $\hat\theta$, a fixed value of $\theta$.
 
 \vspace{1cm}
\noindent{\bf A toy example: the Poisson model}\\Suppose $Y\sim \text{Po}(\lambda)$, and a priori $\lambda\sim\text{Exp}(1)$. 
The marginal likelihood in case is 
\begin{equation}
p(y) = \int_0^\infty \left(e^{-\lambda(n+1)}\lambda^{\sum_{i=1}^n}y_i/\prod_{i=1}^ny_i!\right)\,d\lambda = \frac{\Gamma(\sum_i y_i+1)}{\prod_iy_i (n+1)^{\sum_iy +1}}\label{eq:marglik}
\end{equation}

\begin{figure}
\includegraphics[scale=0.5,angle=-90,keepaspectratio]{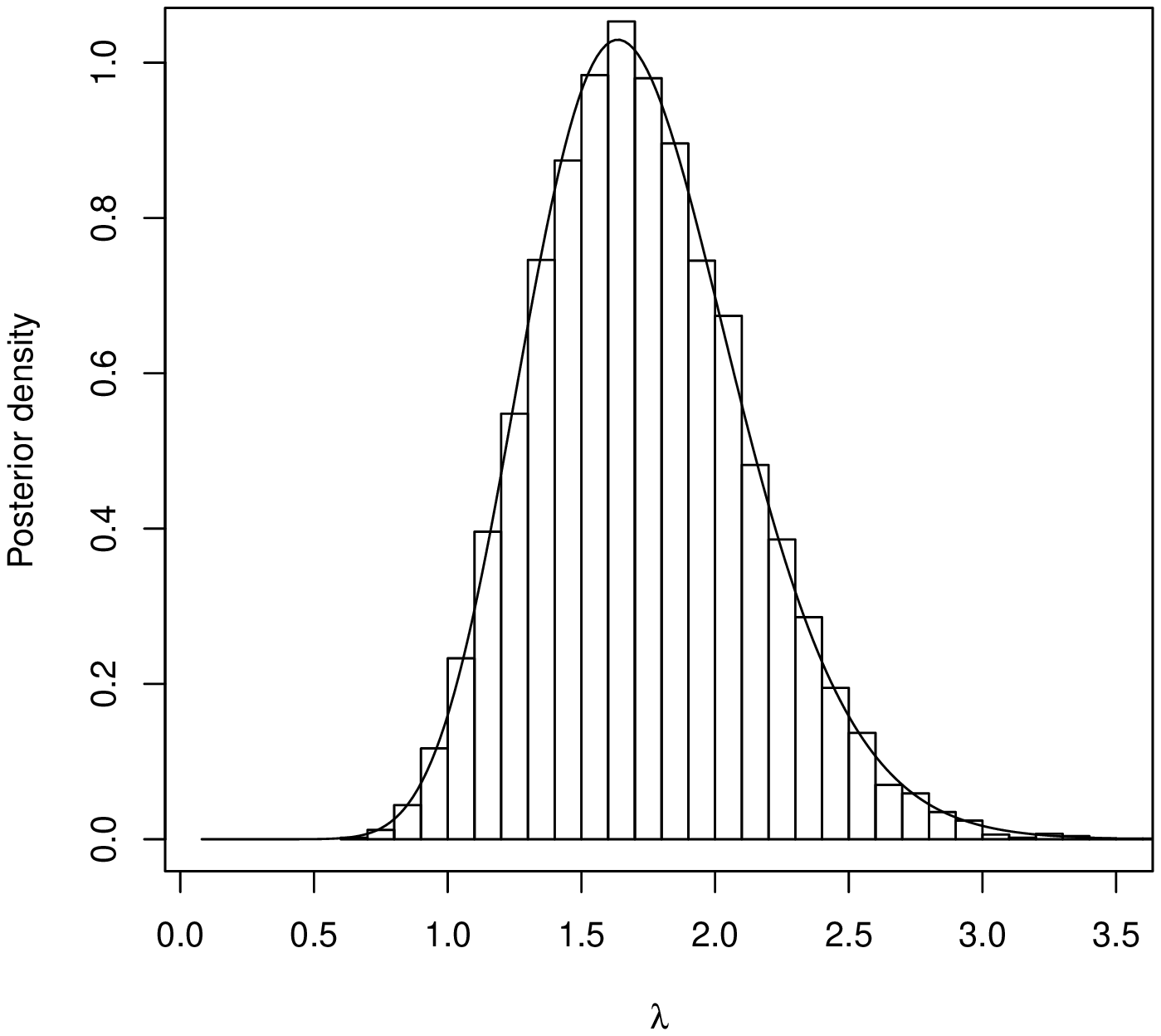} 
\includegraphics[scale=0.5,angle=-90,keepaspectratio]{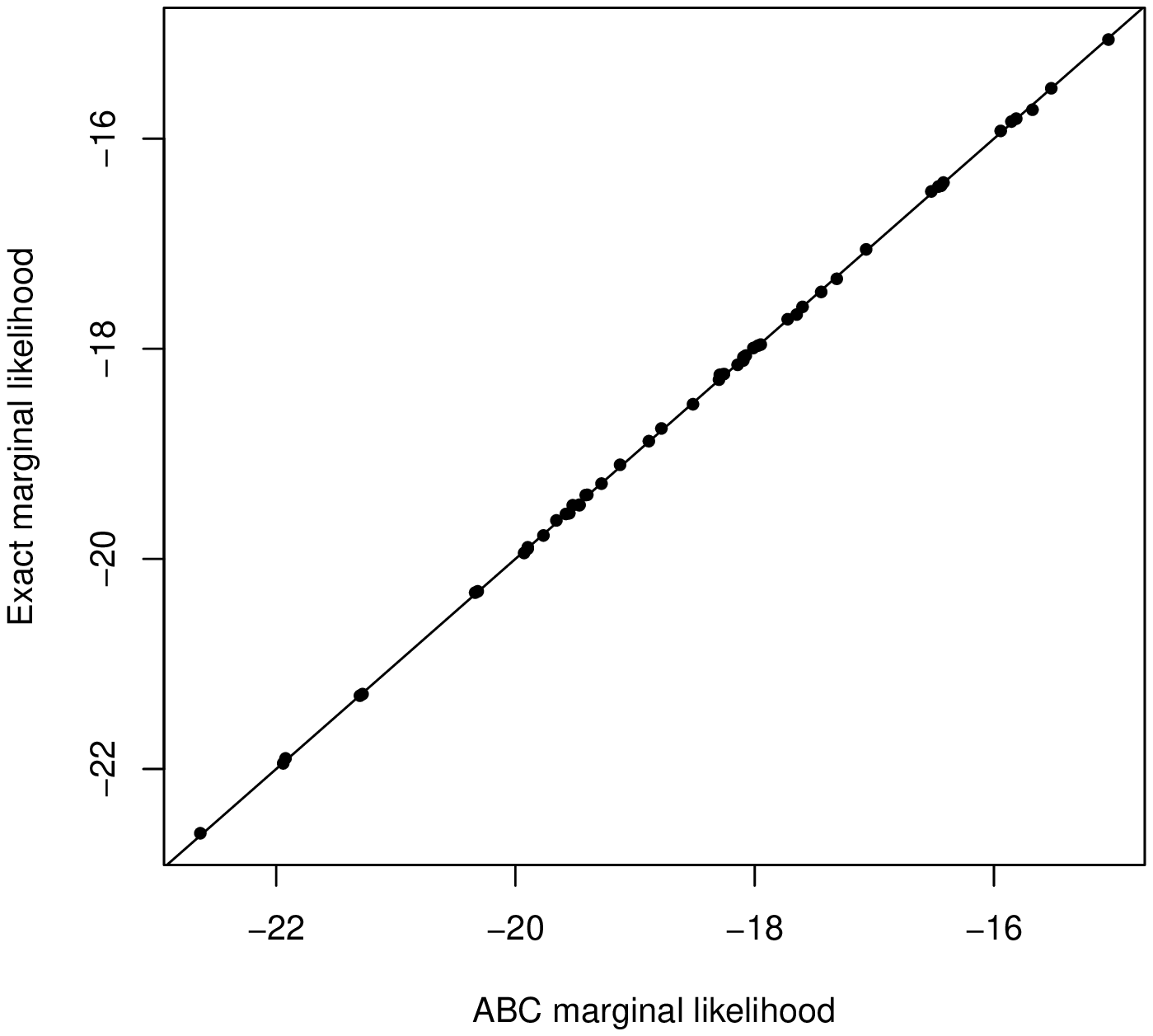}\\
\caption{Left: ABC (histogram) against the exact posterior (line). Right: scatter plot of the log-marginal likelihoods obtained with ABC versus the exact log-marginal likelihoods in 50 random samples, each of size $10$.}
\label{fig:fig}
\end{figure}

As a numerical example, consider $y = (2, 3, 1, 1, 2, 1, 3, 1, 3, 1)$, which are realisations of $10$ random draws from $\text{Po}(2)$ distribution. Figure~\ref{fig:fig} shows on the left side the histogram of the ABC posterior against the exact posterior (solid line). The ABC posterior is approximated by $1\times 10^4$ final samples with $\epsilon=0.001$, where $\rho(\cdot)$ is the Euclidean distance among the total number of counts.
The right side of Figure~\ref{fig:fig} shows the scatter plot of the logarithm of the marginal likelihood obtained from ABC against the logarithm of the exact marginal likelihood \eqref{eq:marglik}, in 50 random samples of size $10$ from $\text{Po}(2)$. The approximate marginal likelihoods obtained from ABC with the sufficient statistic and the exact marginal likelihoods are virtually indistinguishable.

\section{Conclusion}
Obviously, in realistic scenarios sufficient summary statistics are unavailable. However, if we have a set of judiciously chosen summary statistics which give provably valid inference on the parameters of interest, then the idea can still be usefully applied. The more close to sufficiency is the summary statistic the more close to the exact value is the proposed approximation.

\section*{Acknowledgement}
This work was presented at the BayesComp 2018 conference (26--29 March, Barcelona) and is partially supported by University of Padova (Progetti di Ricerca di Ateneo 2015, \texttt{CPDA153257}) and by PRIN 2015 (grant \texttt{2015EASZFS\_003}).

\bibliographystyle{apa}
\bibliography{biblio}

\begin{thebibliography}{}

\bibitem[\protect\astroncite{Barnes et~al.}{2012}]{barnes2012}
Barnes, C.~P., Filippi, S., Stumpf, M. P.~H., and Thorne, T. (2012).
\newblock Considerate approaches to constructing summary statistics for {ABC}
  model selection.
\newblock {\em Statistics and Computing}, 22(6):1181--1197.

\bibitem[\protect\astroncite{Grelaud et~al.}{2009}]{grelaud2009abc}
Grelaud, A., Robert, C.~P., Marin, J.-M., Rodolphe, F., Taly, J.-F., et~al.
  (2009).
\newblock {ABC likelihood-free methods for model choice in Gibbs random
  fields}.
\newblock {\em Bayesian Analysis}, 4:317--335.

\bibitem[\protect\astroncite{Marin et~al.}{2014}]{marin2014relevant}
Marin, J.-M., Pillai, N.~S., Robert, C.~P., and Rousseau, J. (2014).
\newblock Relevant statistics for bayesian model choice.
\newblock {\em Journal of the Royal Statistical Society: Series B (Statistical
  Methodology)}, 76:833--859.

\bibitem[\protect\astroncite{Marin et~al.}{2012}]{marin2012}
Marin, J.-M., Pudlo, P., Robert, C.~P., and Ryder, R.~J. (2012).
\newblock {Approximate Bayesian computational methods}.
\newblock {\em Statistics and Computing}, 22:1167--1180.

\bibitem[\protect\astroncite{Robert et~al.}{2011}]{robert2011lack}
Robert, C.~P., Cornuet, J.-M., Marin, J.-M., and Pillai, N.~S. (2011).
\newblock Lack of confidence in approximate bayesian computation model choice.
\newblock {\em Proceedings of the National Academy of Sciences},
  108:15112--15117.

\bibitem[\protect\astroncite{Rubio and Johansen}{2013}]{rubio2013simple}
Rubio, F.~J. and Johansen, A.~M. (2013).
\newblock A simple approach to maximum intractable likelihood estimation.
\newblock {\em Electronic Journal of Statistics}, 7:1632--1654.

\end{thebibliography}

\end{document}